\documentclass[aps,prl,twocolumn,letterpaper,showpacs,preprintnumbers,amsmath,amssymb]{revtex4}

\usepackage{epsfig,amssymb}
\usepackage{graphicx}
\usepackage{dcolumn}
\usepackage{bm}

\begin{document}

\title{``Kohn-Shamification'' of the classical density-functional
theory of inhomogeneous polar molecular liquids with application to liquid hydrogen
chloride} \author{Johannes~Lischner} \author{T.A.~Arias}
\affiliation{Laboratory of Atomic and Solid State Physics, Cornell
University, Ithaca, New York 14853}
\date{\today}

\begin{abstract}
%600 chars (including spaces, not line feeds)
The Gordian knot of density-functional theories for classical
molecular liquids remains finding an accurate free-energy functional
in terms of the densities of the atomic sites of the molecules.
Following Kohn and Sham, we show how to solve this problem by
considering noninteracting molecules in a set of effective potentials.
This shift in perspective leads to an accurate and computationally
tractable description in terms of simple three-dimensional functions.
We also treat both the linear- and saturation- dielectric responses of
polar systems, presenting liquid hydrogen chloride as a case study.

\end{abstract}
\maketitle

{\em Introduction ---} The importance of inhomogeneous polar molecular
liquids, above all water, in the physical sciences can hardly be
overstated: as solvents they are ubiquitous in soft condensed matter
physics, biophysics, nanophysics, and chemical physics.  The
associated phenomena include hydrophobic interactions
\cite{ChandlerHydrophobic, HydrophobicReview}, protein folding
\cite{ProteinBerne,ProteinTarek}, the behavior of colloidal
suspensions \cite{WennerstromColloid}, and phase transitions of
confined liquids \cite{ConfinedHummer,ConfinedLeng}.  Due to the
complex interplay of hydrogen bonding, long-range polar interactions,
and short-range excluded volume effects, developing a tractable physical theory
to describe the solvent in these systems remains a challenge
\cite{SimpleLiquids}.

Despite the importance, inherent interest, and extensive
experimental study of polar liquids, most existing theories of the
inhomogeneous molecular liquid either lack the accuracy to describe
the above phenomena in a quantitatively satisfying way or become
computationally prohibitive when applied to such complex problems.
The description of liquids from first principles via \emph{ab initio}
methods may be accurate, but can only be applied to relatively small
systems \cite{Galli}. Even {\em classical} molecular dynamics requires
rather long simulation times to sample phase-space sufficiently to
extract meaningful thermodynamic averages \cite{Gunsteren}.  
This latter approach also suffers from the well-recognized difficulty of designing
potentials to describe hydrogen-bonded liquids accurately
\cite{HalgrenDamm}.

An alternative approach is to start with the quantum mechanical
free-energy functional for both electrons and nuclei and, by
``integrating out'' the electrons, to construct a density functional in
terms of atomic site densities alone. These ``classical'' density-functional
theories, which have been successfully applied to the study of simple
liquids \cite{SaamEbner,AshcroftCurtin2}, provide a description of
inhomogeneous physical systems that is founded on a number of exact
theorems \cite{Mermin,Ashcroft}.  To apply this approach to the study
of simple liquids, a hard
sphere reference system is usually augmented by terms that capture
weak long-range attractive forces \cite{Weeks}.  Unfortunately, for
most liquids of interest, a hard sphere
reference system is a poor starting point because of the strong anisotropic
short-range interactions arising from the molecular structure and
effects such as hydrogen bonding.

To remedy this, Chandler and coworkers
\cite{Chandler1, Chandler2, ChandlerWater} introduced a
density-functional theory for molecular liquids in terms of a set of
densities, one for each ``interaction site'' on the molecule
(typically atomic centers).  However, the construction of accurate
free-energy functionals in such theories is challenging due to the
``inversion problem'', the difficulty of using {\em only} atomic site
densities to express the entropy associated with the geometric
structure of the molecules.  Below, we show how this inversion problem
can be overcome by a Kohn-Sham-like change of variables from site
densities to effective potentials and how the resulting functionals
are both computationally tractable and can capture the basic
underlying physics of molecular liquids, including dielectric screening effects.

{\em Kohn-Shamification ---}
The grand free energy $\Omega^{(ni)}$ of a noninteracting gas of
molecules is well-known as a functional of the {\em relative
potentials}
$\psi_{\alpha}(\mathbf{r})\equiv\phi_{\alpha}(\mathbf{r})-\mu_{\alpha}$,
where $\phi_{\alpha}(\mathbf{r})$ is the site-dependent external
potential and $\mu_{\alpha}$ is a site-specific chemical potential,
\begin{eqnarray}
\Omega^{(ni)}=-k_{B}Tn_{r} 
\int d^{3M}r\;s(\{\mathbf{r}_\alpha\}) 
e^{-\beta\sum_{\alpha=1}^{M}\psi_{\alpha}(\mathbf{r}_{\alpha})}.
\label{freeEni}
\end{eqnarray}
Here, $n_{r}$ is the reference density at vanishing chemical
potentials, $M$ is the number of interaction sites on the molecule,
and $s(\{\mathbf{r}_\alpha\})$, which describes the geometry of the
molecule, is the intra-molecular distribution function.  For rigid
molecules (zero internal energy), the entropy of the atomic site
densities is easily extracted from $\Omega^{(ni)}$.  Thus, to
construct an {\em exact} density-functional for the entropy, even for
the {\em interacting} system, one only need express $\Omega^{(ni)}$ as
a functional of the site densities.  More generally, $\Omega^{(ni)}$
is a key part of the interacting functional directly analogous to the
noninteracting kinetic energy functional $T_s[n]$ of Kohn and Sham
\cite{KohnSham}.  Unfortunately, (\ref{freeEni}) only gives
$\Omega^{(ni)}$ as a functional of the $\psi_\alpha$.  To express
$\Omega^{(ni)}$ in terms of site densities requires solution of
$n^{(ni)}_{\alpha}(\mathbf{r})={\delta\Omega^{(ni)}}/{\delta\psi_{\alpha}(\mathbf{r})}$
as a set of $M$ coupled integral equations for the
$\psi_\alpha(\mathbf{r})$ in terms of the
$n^{(ni)}_\alpha(\mathbf{r})$.  This constitutes the ``inversion
problem'' described above.

Chandler and coworkers have solved this inversion problem analytically
in terms of an infinite continued fraction of convolutions for the
{\em di}atomic case {\em only} \cite{Chandler2}.  The unwieldiness of
this formal solution and its limitation to diatomics, however, has led
actual calculations to be performed in the {\em united-atom
approximation}, where, for the noninteracting part of the functional,
all of the sites of a molecule are assumed to coincide at a single
point \cite{ChandlerWater}.  Despite its crudeness, the resulting
theory predicts quite well the correct stable ice structure at nearly
the correct density \cite{ChandlerWater}.  However, by uniting the
sites, this approximation cannot capture the dielectric response of an
ideal gas of polar molecules and so provides a poor starting point for
the study of dielectric effects.

The key observation which allows us to overcome the inversion problem
is that the free energy of a noninteracting molecular system, being a
very complicated functional of the site densities, is a very simple
functional of the relative potentials.  This parallels the situation
in electronic density-functional theory, where there is no known
accurate functional for $T_s[n]$, the noninteracting (kinetic) energy,
in terms of the density, but where this energy is easily written {\em
exactly} in terms of single-particle orbitals.  It was the change in
variables from the density to the single-particle orbitals of
fictitious noninteracting particles, as proposed by Kohn and Sham
\cite{KohnSham}, which allowed for the construction of accurate
density functionals for the interacting electron gas.  In the present
case, the simplicity of evaluation of the grand free energy of
noninteracting molecules as a functional of the relative potentials
suggests an analogous change of variables, now from the site densities
to a set of effective relative potentials in which fictitious
noninteracting molecules move.

Mathematically, a pair of Legendre transformations achieves this
change of variables.  Thermodynamically, the Legendre
transform $\Omega^{(ni)}-\sum\int dr\,\psi_{\alpha}n_{\alpha}$ equals
the intrinsic Helmholtz free energy of the noninteracting system.
Adding the internal energy $U$ due to inter-molecular
interactions and the contribution of the {\em physical} external
potentials $\phi_{\alpha}$ yields the full Helmholtz free energy of
the {\em interacting} system. Finally, the second Legendre transformation
subtracts $\sum\int dr\, n_{\alpha}\mu_{\alpha}$ to form the full {\em interacting} grand
free energy $\Omega$ as a functional of the $\psi_\alpha$,
\begin{equation}
\Omega = \Omega^{(ni)}-
\sum_{\alpha=1}^{M}\int d^{3}r \left(
\psi_{\alpha}(\mathbf{r})-\phi_{\alpha}(\mathbf{r})+\mu_{\alpha}\right)
n_{\alpha}(\mathbf{r}) +U[\mathbf{n}],
\end{equation}
where $\bm{n}=\{n_{1}(\bm{r}),...,n_{M}(\bm{r})\}$, the set of
densities, is explicitly a functional of the effective relative
potentials $\mathbf{\Psi}=\{\psi_{1}(\bm{r}),...,\psi_{M}(\bm{r})\}$
via
$n_{\alpha}(\mathbf{r}) = n^{(ni)}_{\alpha}[\mathbf{\Psi}](\mathbf{r}) \equiv
{\delta\Omega^{(ni)}}/{\delta\psi_{\alpha}(\mathbf{r})}$.
The effective relative potentials that
minimize $\Omega[\mathbf{\Psi}]$ then determine the equilibrium site densities and
allow for the calculation of the various equilibrium properties of the
liquid.

{\em Construction of approximate functionals ---} As usual with
density-functional theories, the construction of the internal 
energy $U$ is difficult.  Again, we follow the lead of Kohn and
Sham and construct a free-energy functional that reproduces 
established results for the homogeneous phase in the limit of
vanishing external fields.  The Ornstein-Zernike (OZ) equation gives
information about the analytic structure of $U$, in particular it
gives $\partial^2 U$, the Hessian of $U$ with respect to the densities, as the difference between the inverses of the
full and noninteracting correlation function matrices \cite{Henderson}.  In
the special case of the homogeneous phase, translational invariance
then turns the OZ equation into a simple matrix equation in Fourier space, with
one component for each field.

From now on, we will limit the discussion to a special class of
liquids, which includes all diatomics and the liquid of most interest,
water. The liquids in this class have the property that the lowest-order term in the long-wavelength expansion of the Hessian of
$U[\mathbf{n}]$ has the form
\begin{equation}
K_{\alpha\gamma}(\mathbf{k})=\left(\frac{\epsilon}{\epsilon-1}-
\frac{\epsilon^{(ni)}}{\epsilon^{(ni)}-1}\right)
\frac{4\pi}{\mathbf{k}^{2}}q_{\alpha}q_{\gamma},
\label{K}
\end{equation}
where $q_{\alpha}$ are the partial site charges, $\epsilon$ is the
macroscopic dielectric constant and $\epsilon^{(ni)}$ is the
dielectric constant of a system with intra-molecular correlations
only.  This result may be derived by expanding the interacting and
noninteracting correlation functions as $E+F k^{2} + \ldots\,$, where
there are no linear terms in $k$ by rotational invariance, and $E$ and $F$
are matrices of coefficients, with one combination of coefficients from
$F$ giving the dielectric constant.  In
the cases of molecules described either by two sites or by three sites
with a mirror plane symmetry, one can show that the {\em only} combination of the
coefficients in $F$ which enters the leading-order term in the
OZ equation corresponds {\em exactly} to the bulk
dielectric constant.  All liquids composed of such molecules thus have the
property that the long-wavelength dielectric response can be built
into the free-energy functional without {\em any} knowledge of the
long-wavelength limit of the experimental correlation functions (other than $\epsilon$).

In general, we may view the internal energy $U$ as expanded in a
power series about the uniform liquid, with all higher-order terms
gathered into an excess part, $F^{ex}$, which plays exactly the same role
as the exchange-correlation functional of Kohn-Sham theory: it
ensures proper bulk thermodynamic behavior and will ultimately be
treated in some approximate way.  Below, the constant term in the power
series for $U$ also will be
handled as part of $F^{ex}$.  Finally, the linear term vanishes in the
uniform equilibrium state of the system, and the quadratic term is given by
the Hessian, which (\ref{K}) gives in the long-wavelength
limit.  Putting this all together gives
\begin{eqnarray}
U[\mathbf{n}]=\frac{1}{2}\sum_{\alpha,\gamma=1}^{M}\int
d^{3}r \int d^{3}r'\; n_{\alpha}(\mathbf{r})\left\{
K_{\alpha\gamma}(\mathbf{r},\mathbf{r'})+ \right. \nonumber\\
\left. C_{\alpha\gamma}(\mathbf{r},\mathbf{r'})\right\} n_{\gamma}(\mathbf{r'})
+F^{ex}[\mathbf{n}],
\label{U}
\end{eqnarray}
where $C_{\alpha\gamma}$ describes those parts of the Hessian
$\partial^2 U$ which $K$ fails to capture.

To approximate $F^{ex}$, we first note that,
in the case of zero external fields, all densities are equal and the
first quadratic term ($K$) in (\ref{U}) vanishes because of charge
neutrality. Anticipating that the matrix function $C$ will be constructed
to vanish in the long-wavelength limit, $F^{ex}$ then captures all the
internal energy of the uniform phase and can be expressed as
$F^{ex}=Vf^{ex}(n)$, with $V$ being the volume and $n$ the average
molecular density.  Due to the presence of multiple density
fields, generalizing this
expression to the inhomogeneous case is more difficult than for the
analogous exchange-correlation energy in electronic structure
theory. Also, because of the strong correlations induced by excluded
volume effects, {\em purely local} excess functionals fail to describe the
liquid state \cite{CurtinAshcroft}.  We therefore approximate $F^{ex}$ with a simplified {\em ansatz} in the
spirit of weighted density-functional theory \cite{CurtinAshcroft},
but generalized to multiple species,
\begin{equation}
F^{ex}[\bm{n}]=\int
d^{3}r\sum_{i}p_{i}
f^{ex}\left(\sum_{\gamma=1}^{M}b_{\gamma}^{i}\overline{n}_{\gamma}(\mathbf{r})\right),
\label{Fex}
\end{equation} 
where we introduce the weighted densities
$\overline{n}_{\gamma}(\mathbf{r})=\int d^{3}r' (\pi r_{0}^{2})^{-3/2}
\exp(-|\mathbf{r}-\mathbf{r'}|/r_{0}^{2})n_{\gamma}(\mathbf{r}')
$, with $r_{0}$ being a parameter ultimately
fit to the experimental surface tension.  To reduce to
the correct form in the uniform case $p_{i}$ and $b^{i}_{\gamma}$ must
fulfill $\sum_{i}p_{i}=1$ and
$\sum_{\gamma=1}^{M}b_{\gamma}^{i}=1$.

To capture the behavior of the scalar function $f^{ex}(n)$, we use a
polynomial fit to various bulk thermodynamic
conditions.  The condition that $C$ vanishes in the long-wavelength
limit subsequently fixes $p_{i}$ and $b^{i}_{\gamma}$ and ensures
that $C$ will be bandwidth limited and thus amenable to numerical
approximation.  For a given $r_0$, this then completely specifies our
approximation to $F^{ex}$.  Next, relating
$K+C+\partial^2 F^{ex}$ to the density-density correlation
functions through the OZ relation then gives the matrix
function $C$ for a given $r_0$.  Finally, we can determine $r_0$ by
adjustment until calculations of the liquid-vapor interface give the
correct surface tension.

{\em Hydrogen chloride ---} We choose liquid hydrogen chloride as a
model physical system exhibiting hydrogen bonding and for which
detailed experimental data are available, including site-site correlation functions\cite{Soper,Pasquarello}.

Assuming rigid intra-molecular bonds, the distribution function
becomes
$s(\bm{r}_{1},\bm{r}_{2})=\delta(|\bm{r}_{1}-\bm{r}_{2}|-B)/4\pi
B^{2}$, with the gas phase bond length $B=1.275$~\AA\ taken from
experiment \cite{Pasquarello} and $1$ and $2$ referring to hydrogen
and chlorine, respectively.  We then select the partial charges to
yield the experimental gas phase dipole moment \cite{DeLeeuw},
resulting in $q_{1}=-q_{2}=0.171\,e$ with $e$ being the fundamental
charge. Then, we approximate the simple bulk function $f^{ex}(n)$ by a
fourth-order polynomial with coefficients adjusted to (a) reproduce
the bulk modulus of the liquid phase and (b) to allow for coexistence
of stable liquid and vapor phases at the appropriate densities, where
the latter is treated as an ideal gas.  Next, for the excess
free-energy {\em functional}, the requirement that the matrix $C$
vanishes in the long-wavelength limit fixes the integrand in
(\ref{Fex}) to be $[3f^{ex}(\overline{n}_{1}(\mathbf{r}))+
4f^{ex}\left((\overline{n}_{1}(\mathbf{r})+\overline{n}_{2}(\mathbf{r}))/2\right)+
3f^{ex}(\overline{n}_{2}(\mathbf{r}))]/10$, under the assumption that
the constant term (in an expansion in powers of $k^2$) of the inverse
correlation function is proportional to that of the noninteracting
case.  Then, we determine $C$ as described above using the partial
structure factor data for the uniform liquid \cite{Soper} and the
macroscopic dielectric constant \cite{Swenson} as experimental
inputs. Adjusting the smoothing parameter to give the experimental
surface tension \cite{AdlerYaws} yields $r_{0}=4.55$~\AA.

\begin{figure}
\includegraphics[width=7.5cm]{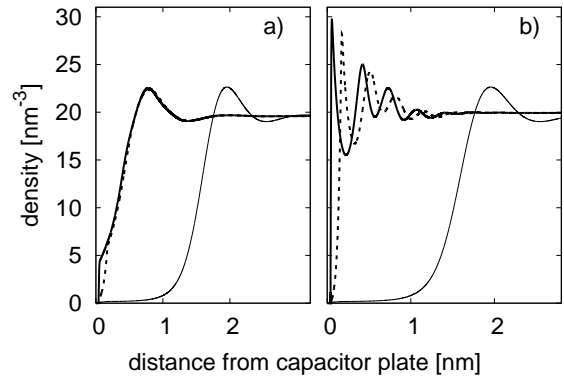}
\caption{Hydrogen- (thick dotted curves) and chlorine- (thick solid
curves) site density versus distance from hard wall in (a) moderate and (b)
high applied field.  Reference results for zero field given in both
panels (thin solid curves).}
\label{SiteDensity}
\end{figure}

{\em Results ---} As a test case, we study the behavior of liquid
hydrogen chloride at T=194 K, P=4.5 bar
(chosen due to availability of experimental correlation
functions \cite{Soper}), which is in the liquid part of the phase diagram somewhere near the
triple point.  Figures~\ref{SiteDensity}(a,b) show the equilibrium
density profiles which our theory predicts in a parallel plate capacitor (described as an
infinite square-well potential for both species) in both moderate and
strong applied fields, respectively.  For comparison, the figure also
shows the zero-field density profiles.

The zero-field profiles exhibit an extended gas
phase region up to a distance of $\sim 1$~nm from the plates. This
``lingering'' gas phase region exists because molecules can minimize
the influence of the repulsive walls by leaving the
system and entering the reservoir at very little free-energy cost.
As soon as a relatively weak external electric displacement $D$ is applied,
however, it becomes favorable for the dipolar molecules to enter the
capacitor, largely destroying the gas phase region and resulting in
an almost rigid shift of the density profiles towards the capacitor
walls for a large range of fields until the non-linear dielectric response regime
is reached.  Figure~\ref{SiteDensity}(a) shows an example of this
behavior for a relatively large field but where the dielectric
response is still fairly linear ($\alpha \equiv \beta p D = 2.68$,
where $p$ is the molecular dipole moment).  A dramatic qualitative
change occurs in the strong field case.  Figure~\ref{SiteDensity}(b), computed at $\alpha=7.38$, shows the typical behavior in the high-field case.  At
large fields, the profile for each species exhibits a sharp peak
followed by strong oscillations, where the peaks and oscillations are separated by one
molecular bond length ($\sim 0.12$~nm) indicating that they are the
result of strong orientational ordering of the molecular dipoles. The
oscillations of the site densities suggest an induced layering of
molecules close to the surface, which results from the reorganization
of the hydrogen-bond network in response to the increasing dipolar
alignment of the molecules. The wavelength of the observed
oscillations is about $\sim 0.3$~nm, corresponding to the length of the
hydrogen bond \cite{SimpleLiquids}.

\begin{figure}
\includegraphics[width=7.5cm]{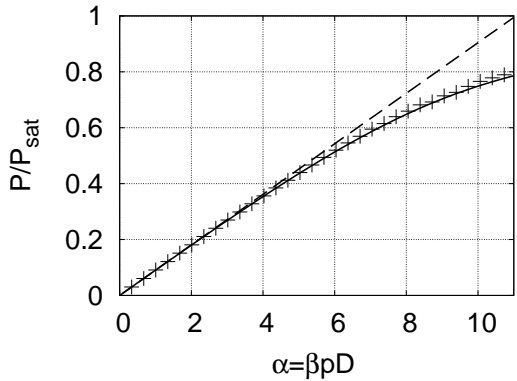}
\caption{Electric saturation fraction ($P/P_{\text{sat}}$) versus
  applied electric displacement ($D$): density-functional results
  (crosses), non-linear electrostatics (solid line), linear response
  (dashed line).}
\label{SurfCharge}
\end{figure}

The equilibrium site densities also determine the induced polarization
$P$. Figure~\ref{SurfCharge} compares the polarization, computed from
our density-functional theory, to the result of self-consistently screened
nonlinear electrostatics, computed by solving for the polarization
$P$ such that $P=P^{(ni)}(D-a_{\epsilon}4\pi P)$, where $P^{(ni)}(E)$
is the response of a gas of noninteracting dipoles in a local field
$E$ and
$a_{\epsilon}\equiv \epsilon/(\epsilon-1)-\epsilon^{(ni)}/(\epsilon^{(ni)}-1)$
ensures that the correct linear response is recovered when $D$ is
weak.  Figure~\ref{SurfCharge} shows that our density-functional theory
not only reproduces the linear regime but also captures
self-consistent saturation effects.
We remark that such results cannot be obtained within the united-atom
approximation or any other approximation that does not take the
intra-molecular bonding geometry into account.

In conclusion, we have shown how a Kohn-Sham-like change of variables
yields a numerically efficient and accurate density-functional theory
for molecular liquids.  The resulting theory has the computational
cost of the problem of a noninteracting gas of molecules in a
self-consistent external potential.  We have shown how to construct a
functional in this approach which captures the coexistence of liquid and
vapor phases, the surface tension between these phases, and, for the
liquid, the bulk mechanical properties, site-site correlation
functions, linear dielectric response, and self-consistent dielectric
saturation effects.  This approach can now be applied to 
liquids in complex environments which can be described as external
potentials, including nanopores, hydrophobic/hydrophilic molecules, and large biosolutes.  Without
modification, the current approach is ready to be developed for the liquid of
greatest scientific interest: water.

\end{document}